\documentclass[final]{raa}
\usepackage{graphicx,times}
\usepackage{natbib}

\providecommand{\bm}[1]{\mbox{\boldmath$#1$\unboldmath}}

\begin{document}

   \title{GRB Jet Beaming Angle Statistics
}

 \volnopage{ {\bf 2009} Vol.\ {\bf 9} No. {\bf XX}, 000--000}
   \setcounter{page}{1}

   \author{Y. Gao
      \inst{}
   \and Z.G. Dai
      \inst{}
   }

   \institute{Department of Astronomy, Nanjing University,
             Nanjing 210093, China\\
\vs \no
   {\small Received [year] [month] [day]; accepted [year] [month] [day] }
}

\abstract{ Existing theory and models suggest that a Type I (merger)
GRB should have a larger jet beaming angle than a Type II
(collapsar) GRB, but so far no statistical evidence is available to
support this suggestion. In this paper, we obtain a sample of 37
beaming angles and calculate the probability that this is true. A
correction is also devised to account for the scarcity of Type I
GRBs in our sample. The probability is calculated to be 83\% without
the correction and 71\% with it.
  }

   \authorrunning{Y. Gao \& Z.G. Dai }            
   \titlerunning{GRB Jet Beaming Angle Statistics }  
   \maketitle


%
%
\section{Introduction}           
\label{sect:intro}

There are two intrinsically different phenomena that give rise to
gamma-ray bursts (GRBs). One is the merging of two compact objects,
such as that of a neutron star and a black hole; the other is the
core collapse of a massive star ``collapsar"), such as the birth of
a Type Ib/c Supernova. According to a classification scheme
developed recently (Zhang, 2007 and Zhang et al, 2009), a GRB
obtained from the former channel is defined as a Type I GRB, while
one from the latter would be classified as a Type II GRB. This
classification scheme, though more intrinsic than the classical
short/hard vs long/soft categories, is not easy to carry out at the
present stage, since many different criteria need to be applied in
order to determine the progenitor of a GRB, few of which are
decisive. One of these criteria is that Type I GRBs are usually
short/hard (
\begin{math}
T_{90}\leq2\,{\rm s}
\end{math}
), while Type II GRBs are usually long/soft (
\begin{math}
T_{90}>2\,{\rm s}
\end{math}
). This criteria, though supported by many individual cases, is not
decisive, as any short GRB could theoretically have been a long GRB
had it occurred at a high enough redshift. Of the criteria that are
decisive, such as whether or not a GRB emanates gravitational
radiation, most are impractical and the remainder is not applicable
to a great majority of GRBs at the present stage.

GRBs eject their energy in the form of jets. These jets have already
been modeled, and their beaming angles (also called opening angles)
can be calculated from observed physical data (Sari et al, 1999).
The size of the beaming angle, according to the models, is highly
dependent on the nature of the GRB progenitor: theoretically, Type
II GRBs should have smaller beaming angles than Type I GRBs, due to
the collimation effect of the stellar wind hugging collapsars.
However, this dependency has not yet been supported statistically,
possibly due to the aggravatingly small sample of Type I GRB jet
beaming angles available, hence this paper, which specifically seeks
to find statistical evidence of this  theoretical relationship.

This paper is divided into 5 sections, of which this introduction is
the first. In the second, we will list the relevant observational
data and use it to obtain the jet beaming angles of 37 GRBs. The
third section will cover data reduction and statistical analysis.
The results and discussion of the analysis will be presented in the
fourth section, and the entire paper will be summarized in the fifth
section, which is the conclusion.

\section{Collected Data}
\label{sect:Obs}

According to the literature (Sari et al, 1999), the jet beaming
angle of a GRB could be obtained by Eq.(1):
\begin{equation}
\theta_{j}=0.057(\frac{t_{j}}{1\,{\rm day}})^{\frac{3}{8}}
(\frac{1+z}{2})^{-\frac{3}{8}} (\frac{E_{iso}(\gamma)}{10^{53}\,{\rm
erg}})^{-\frac{1}{8}}
(\frac{\eta_{\gamma}}{0.2})^{\frac{1}{8}}(\frac{n}{0.1\,{\rm
cm}^{-3}})^{\frac{1}{8}},
\end{equation}
where $\theta_{j}$ is the jet beaming angle mentioned above, $t_{j}$
is the time of the jet break, $z$ is the redshift at which the GRB
took place, $E_{iso}(\gamma)$ is the isotropic energy released by
the burst, $\eta_{\gamma}$ is the gamma-ray radiative efficiency of
the jet and $n$ is the density of the interstellar medium. $t_{j}$,
$z$ and $E_{iso}(\gamma)$ are observable parameters, shown in Table
1 (Appendix A). $\eta_{\gamma}$ and $n$, however, are not
observable, but thankfully do not have a very significant effect on
the value of $\theta_{j}$. Therefore, we set $\eta_{\gamma}=0.2$ and
$n=0.1\,{\rm cm}^{-3}$. The jet beaming angles obtained from these
data are presented in Appendix A and in Fig. 1 below. In Fig. 1,
the bars filled in black, stripes and blue correspond to Sample I,
Sample II and unclassified jet beaming angles (defined below)
respectively.

   \begin{figure}[h!!!]
   \centering
   \includegraphics[width=9.0cm, angle=0]{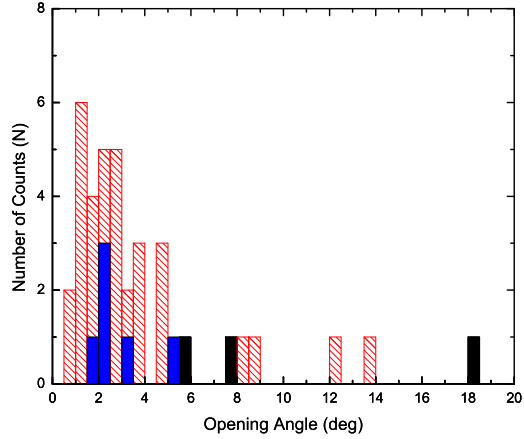}
   \begin{minipage}[]{85mm}
   \caption{ Jet beaming angles: complete sample. Sample I is shown
    in black filling, while Sample II is shown with striped filling.
    Filled in blue are those GRBs that have not been positively identified
    as either Sample I or Sample II. The horizontal axis is $\theta_{j}$
    in degrees, while the vertical axis the number of opening angles that
    are of the specified magnitude.} \end{minipage}
   \label{Fig1}
   \end{figure}

\section{Data reduction}
\label{sect:data}

As shown in Appendix A, we have only managed to confirm 2 Type I
bursts and 13 Type II bursts among the GRBs whose parameters we
obtained.

Of the remaining GRBs, one is identified to be a short burst, while
21 of the others are known to be long. According to experience,
short GRBs are usually of a Type I origin, while long bursts are
usually of Type II. Therefore, we incorporate the short GRBs into
the Type I sample, and the long GRBs into the Type II sample. It is
hoped that, statistically, this will serve the purpose of giving a
larger sample while maintaining sample integrity.

Six bursts, namely GRB970828, GRB990510, GRB990705, GRB991216,
GRB000301C and GRB010222 cannot be identified as either Type I/II or
long/short bursts, and therefore cannot be used in the statistics.
These bursts are represented by the columns filled in blue in Fig.
1.

Thus, we obtain a ``Type I + short" sample, noted as ``Sample I" for
the rest of the paper, and a ``Type II + long" sample, hereby noted
as ``Sample II". Their jet beaming angles are presented in Fig. 1
(black and striped filling respectively). A primary contributing
factor to the scarcity of Sample I data is the fact that $t_{j}$ is
hard to determine in short bursts, rendering Eq. (1)
unapplicable. Of the five bursts known to be of Type I origin (Zhang
et al, 2009), we obtained the opening angles for two of them (see
Appendix A for details). Of the remaining three (GRB050509B,
GRB050724 and GRB061006), the jet break time of GRB050509B and
GRB061006 cannot be found, while only a lower limit constraint could
be placed on the jet break time of GRB050724. The problem of whether
lower limit constraints could be used in the sample is discussed
further in section 4. Of the short/hard bursts listed in Zhang et
al, 2009, the opening angle of only one could be ascertained. Since
GRB080503, the latest short/hard burst mentioned in Zhang et al,
2009, five bursts (GRB080702A, GRB080905A, GRB080919, GRB081024 and
GRB090510) have been observed according to GCN reports that fall
indisputably in the short/hard category. However, no jet break time
has been found for any of them.

From Fig. 1, it should be quite obvious that Sample I data have a
greater arithmetic mean in comparison to Sample II data. This is
 indeed the case, as Sample I has a mean value of 10.42 degrees, while
the mean of Sample II is only 3.42 degrees. This is supportive of
the statement that Type I GRBs have a larger beaming angle than Type
II GRBs. However, the fact that there are 4 of the Sample II data
that are larger than 2 of the 3 Sample II data leaves plenty of room
for argument. Therefore, statistical analysis based on the data that
lead to a quantitative result on the validity of the statement above
is required. In order to obtain such a quantitative result,
statistical fitting must be carried out. Many papers consider the
Gaussian distribution under similar circumstances (for instance:
Zhang and Meszaros, 2002), and therefore we will fit the data using
the Gaussian distribution. Whether opening angles really do follow
Gaussian distributions is indeed rather problematic: this
 topic will be discussed further below.

Assuming that elements taken from Samples I and II are random
variables that have Gaussian distributions ($N(\mu,\sigma^{2})$), we
take their probability densities to be

\begin{equation}
N(\theta_{j})=\frac{1}{\sigma_{i}\sqrt{2\pi}}\exp[-\frac{1}{2}(\frac{\theta_{j}-\mu_{i}}{\sigma_{i}})^{2}],
~~~~~(i=1,2),
\end{equation}
respectively. Here, $i=1$ corresponds to Sample I, while $i=2$
corresponds to Sample II. $\mu_{i}$ and $\sigma_{i}$ are the means
and standard variations of the two samples respectively, and are
hereby defined mathematically as $\mu_{i}=\sum\limits_{k}{X_{ki}}$
and $\sigma_{i}^2=\sum\limits_{k}{{(\mu_{i}-X_{ki})}^2}$, where
$X_{ki}$ is the "k"th datum from Sample i.

Taking the sample mean and sample standard deviation as $\mu$ and
$\sigma$ respectively for both samples ($\mu_{1}=10.42$,
$\mu_{2}=3.42$, $\sigma_{1}^{2}=46.24$, $\sigma_{2}^{2}=9.07$), we
obtain best-fit curves as shown in Figs. 2 and 3, where they have
been superimposed on their respective sample distributions for
comparison. For the rest of this paper, these best-fit distributions
for Sample I and Sample II will be termed Fit I and Fit II
respectively. A Kolmogorov-Smirnov test has been preformed for Fit
II, resulting in $D_{34}=0.125$, in other words, an acceptance level
of $1-\alpha\approx65\%$. Fit I is expected to have a very low
acceptance level, which we did not calculate but are certain that it
is (possibly much) smaller than 30\%, which is one of the reasons
why we decided to devise the correction below.

   \begin{figure}[h!!!]
   \centering
   \includegraphics[width=9.0cm, angle=0]{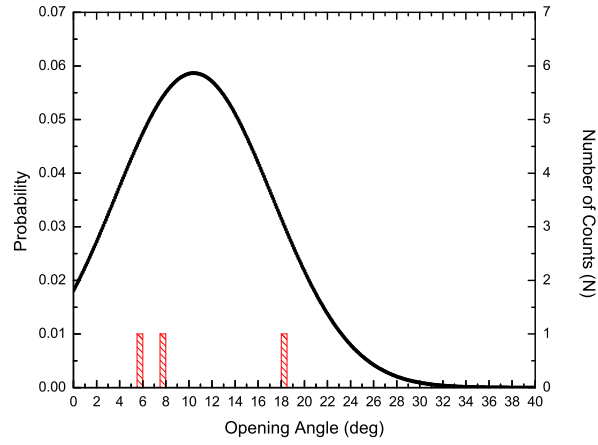}
   \begin{minipage}[]{85mm}
   \caption{ Sample I best fit probability distribution curve superimposed on Sample I bar graph. } \end{minipage}
   \label{Fig2}
   \end{figure}

   \begin{figure}[h!!!]
   \centering
   \includegraphics[width=9.0cm, angle=0]{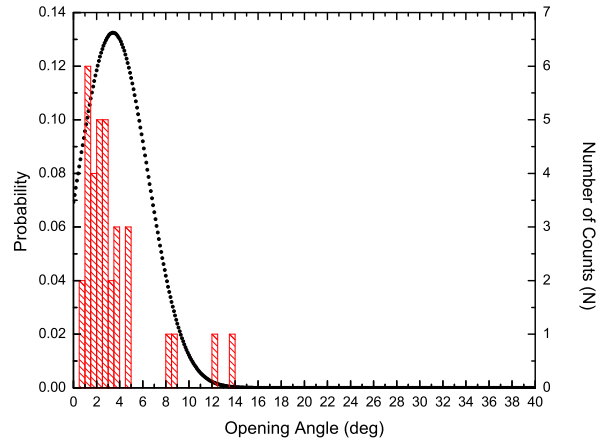}
   \begin{minipage}[]{85mm}
   \caption{ Sample II best fit probability distribution curve superimposed on Sample II bar graph. } \end{minipage}
   \label{Fig3}
   \end{figure}

Next, we calculate the probability that a random variable from Fit I
($\theta_{j1}$) is larger than a random variable from Fit II
($\theta_{j2}$). The random variable $\theta_{j1}-\theta_{j2}$
should have a distribution of

   \begin{figure}[h!!!]
   \centering
   \includegraphics[width=9.0cm, angle=0]{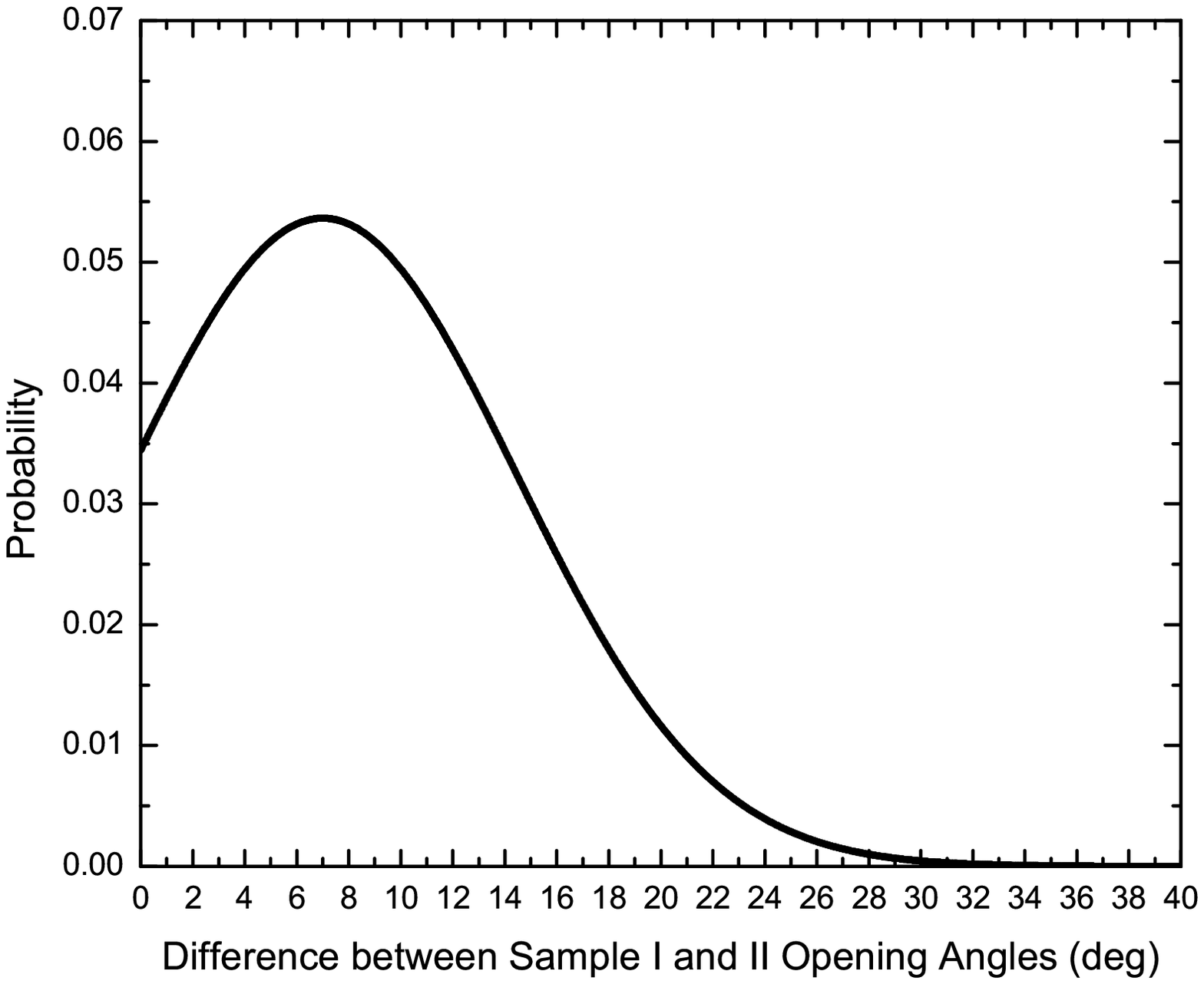}
   \begin{minipage}[]{85mm}
   \caption{ Distribution of random variable
$\theta_{j1}-\theta_{j2}$. This is the probability distribution of
the value of the difference between a random variable from Fit I and
another from Fit II.}
\end{minipage}
   \label{Fig4}
   \end{figure}

\begin{equation}
N(\theta_{j1}-\theta_{j2})=
\frac{1}{\sqrt{\sigma_{1}^{2}+\sigma_{2}^{2}}\sqrt{2\pi}}
\exp[-\frac{1}{2}(\frac{(\theta_{j1}-\theta_{j2})-(\mu_{1}-\mu_{2})}{\sqrt{\sigma_{1}^{2}+\sigma_{2}^{2}}})^{2}],
\end{equation}

This distribution is another normal distribution with a mean of
$(\mu_{1}-\mu_{2})$ and a variance of
$(\sigma_{1}^{2}+\sigma_{2}^{2})$, as shown in Fig. 4. Thus,

\[
P(\theta_{j1}>\theta_{j2})=P(\theta_{j1}-\theta_{j2}>0)=\int_{0}^{+\infty}N(\theta_{j1}-\theta_{j2})d(\theta_{j1}-\theta_{j2}),
\]
i.e.
\begin{equation}
P_{1}(\theta_{j1}>\theta_{j2})=\int_{0}^{+\infty}\frac{1}{\sqrt{\sigma_{1}^{2}+\sigma_{2}^{2}}\sqrt{2\pi}}
exp[-\frac{1}{2}(\frac{x-(\mu_{1}-\mu_{2})}{\sqrt{\sigma_{1}^{2}+\sigma_{2}^{2}}})^{2}]dx.
\end{equation}

Calculating this by means of a Fortran program, we obtain

\begin{equation}
P_{1}(\theta_{j1}>\theta_{j2})=0.83.
\end{equation}

The probability shown in Eq. (5) is the probability that a Type
I jet angle is larger than a Type II jet beaming angle, but only if
Type I and II GRB beaming angles have a distribution strictly the
same as Fit I and Fit II respectively. In the case of Fit II, 34
samples have been taken to make the fit, therefore the difference is
considered negligible. However, for Fit I, which was made with only
three samples, the effects of inconsistency must be considered.

According to Eq. (4), $P(\theta_{j1}>\theta_{j2})$ can be calculated
once $\mu_{1}$ and $\sigma_{1}^{2}$ are given as constants. But the
$\mu_{1}$ and $\sigma_{1}^{2}$ in this equation have probability
distributions of their own, which are reliant on the consistency of
Fit I. They are

\begin{equation}
t(N-1)\sim \frac{(\overline{X}-\mu_{1})\sqrt{N}}{S_{X}},
\end{equation}
and

\begin{equation}
\chi^{2}(N-1)\sim \frac{(N-1)S_{X}^2}{\sigma_{1}^{2}},
\end{equation}
respectively, where $N=3$, the number of Sample I elements. Here, t
and $\chi^{2}$ correspond to a t distribution and a $\chi^{2}$
distribution respectively, $\overline{X}$ is the sample mean,
$S_{X}$ is the sample variance, $\mu_{1}$ and $\sigma_{1}$ are the
theoretical mean and standard variation of Sample I
respectively(shown here as variables).

Integrating Eq. (4) over these probability distributions, taking
$\frac{(\overline{X}-\mu_{1})\sqrt{N}}{S_{X}}$ and
$\frac{(N-1)S_{X}^2}{\sigma_{1}^{2}}$ to be $y$ and $z$
respectively, we write

$~~~~~~~~~~~~~~~~~~~~~~~~~~~~~~~~~~~~~~~~~~~~~~~~~P_{2}(\theta_{j1}>\theta_{j2})$\\
$~~~~~~~~~~~~~~~~~~~~~~~~~~~~~~~~~~~~~~~~~~~~~~~~=\int_{0}^{+\infty}\chi^{2}(z,2)\int_{-\infty}^{+\infty}t(
y,2)\cdot P_{1}dydz$\\
note that $P_{1}$, though given as a constant in Eq.(5), is
dependent on $(\mu_{1}-\mu_{2})$ and
$(\sigma_{1}^{2}+\sigma_{2}^{2})$, which are not constants for the
purpose of the following calculations. Substituting the right hand
side of Eq.(4) for $P_{1}(y,z)$, and equivalent terms in $y$ and $z$
for $\mu_{1}$ and $\sigma_{1}$, we obtain

$~~~~~~~~~~~~~~~~~~~~~~~~~~~~~~~~~~~~~~~~~~~~~~~~~P_{2}(\theta_{j1}>\theta_{j2})$\\
$~~~~~~~~~~~~~~~~~~~~~~~~~~~~~~~~~~~~~~~~~~~~~~~~=\int_{0}^{+\infty}\chi^{2}(z,2)\int_{-\infty}^{+\infty}t(
y,2)\cdot P_{1}(y,z)dydz$
\[
~~~~~~~~~~~~~~~~~~~~=\int_{0}^{+\infty}\chi^{2}(z,2)\int_{-\infty}^{+\infty}t(
y,2)\int_{0}^{+\infty}\frac{1}{\sqrt{g(z)+\sigma_{2}^{2}}\sqrt{2\pi}}
\]

\begin{equation}
\exp[-\frac{1}{2}(\frac{x-(f(y)-\mu_{2})}{\sqrt{g(z)+\sigma_{2}^{2}}})^{2}]dxdydz,
\end{equation}
where
$t(y,2)=\frac{1}{2\sqrt{2}}(1+\frac{y^{2}}{2})^{-\frac{3}{2}}$,
$\chi^{2}(z,2)=\frac{1}{2}exp(-\frac{z}{2})$,
$f(y)=\mu_{1}=\overline{X}-\frac{S_{X}}{\sqrt{3}}y$, and
$g(z)=\sigma_{1}^{2}=(N-1)S_{X}^2/z$.

Again using a Fortran program, we calculate this final result to be

\begin{equation}
P_{2}(\theta_{j1}>\theta_{j2})=0.71.
\end{equation}

\section{Discussion}
\label{sect:analysis}

Theoretically, it would have been more logical to perform a similar
correction on Sample II in conjunction with the one used on Sample
I. However, this would lead to 5-dimensional integration, which
would be far more than what our PC could manage in a short period of
time (the 3-D integration for $P_{2}$ took 10 minutes at double
precision, with a step length of 0.03 for $x$,$y$ and $z$). Instead,
we performed the 3-D integration correction on Samples I and II
respectively. Taking into account the first 4 digits after the
decimal point, $P_{1}$ is calculated to be 0.8272. This merely
decreases to 0.8254 after the application of the correction to
Sample II, while it decreases to $P_{2}=0.7118$ after application of
the correction to Sample I. From these numbers, we find that
negligence of the Sample II correction induces an error of only the
magnitude of $10^{-3}$ (i.e. the third digit after the decimal
point), and thus find it safe to retain the first 2 digits without
performing the correction to Sample II, hence the number of digits
used in the presentation of our results (Eqs. (5) and (9)).

With an acceptance level of only 65\% for even Fit II, it is indeed
debatable whether a Gaussian fit is appropriate for the Samples.
However, due to the fact that there does not exist an established
fit for opening angles, a Gaussian fit seems to be the natural
choice. Further research may experiment on other parametrical fits
that may yield better acceptance levels.

It has been proposed that GRBs which would fall into the Sample I
category, but have only minimum constraints for their jet break
times (such as GRB050724) could be used in the statistics as well.
In doing so, a minimum value could be calculated for both $P_{1}$
and $P_{2}$, on the basis of a larger sample. However, this could
lead to complexities, since there are GRBs which would fall into the
Sample II category that have only minimum constraints for their jet
break times too. In this paper, we wish to avoid such complexities,
and therefore use only exact data.

The reader might want to notice that there are several problems that
have not been taken into account. Firstly, all data on
$E_{iso}(\gamma)$, $z$ and $t_{j}$ in this paper have been collected
from the literature. While this should cause no problems for $z$ and
$E_{iso}(\gamma)$, it could be somewhat problematic for $t_{j}$,
since the time of the jet break is different for different bands at
which the specific observation was made. Throughout this paper, this
matter has been treated indiscriminately.Also, the data, as shown in
Table 1 (Appendix A), are presented with a certain deviation in the
literature. In other words, it is not precise. Even more prominent
is the crisis that if we take Fit I to be exactly the distribution
of Type I GRB jet beaming angles, a significant number of Type I jet
beaming angles will be smaller than zero. These problems, throughout
the paper, have been treated as insignificant details, hereby
submitted to future scrutiny by the reader.

\section{Conclusions}
\label{sect:analysis}

In this paper, we have obtained a Sample I of 3 jet beaming angles
and a Sample II of 34 jet beaming angles. These two samples are
expected to be representative of Type I and Type II GRBs
respectively. After that, normal (Gaussian) distributions have been
fitted to the samples, resulting in Fit I and Fit II. Taking these
fits to be representative of Type I and Type II GRBs, we then
proceed to calculate the probability that a random variable taken
from Fit I is larger than another taken from Fit II, thereby
deriving the probability that a Type I GRB has a larger opening
angle than a Type II GRB. Taking into account the uncertainty caused
by the very small Sample I, we then devise a correction for the
probability stated above, and reach a probability that is much lower
but still significantly high nevertheless.

If we take Sample I to be a sample perfectly representative of Type
I GRBs, then it could be said, with an 83\% degree of certainty,
that Type I GRBs have larger jet beaming angles than Type II GRBs.
However, this 83\% drops to 71\% once we take into account the
uncertainty caused by our small sample in Sample I. In either case,
it could be justifiably concluded, from our current samples, that
Type I GRBs generally have a larger beaming angle in comparison to
Type II GRBs.

Our results are supportive of current models and theory. Type I GRBs
in general have larger beaming angles in comparison to Type II GRBs,
with a fairly high degree of certainty, though by far not high
enough to become a decisive criterion as to whether a GRB is of Type
I or Type II origin (so far).

It is possible that with a larger Sample I, higher levels of
certainty could be reached. Therefore, further observations that
yield data concerning Sample I jet beaming angles (preferably Type I
GRB jet beaming angles) are required for progress. It is also
possible that with a large enough Sample I, the correction methods
used in this paper (i.e. the correction for Sample I) will become
obsolete, however, given the current rate at which Sample I opening
angles are being derived, it seems unlikely that this would be the
case in the near future. Lastly, it would be desirable to find a
parametrical fit which could be supported theoretically, and which
yields a higher level of acceptance than the Gaussian.

\normalem
\begin{acknowledgements}
Many thanks to our research group at the Department of Astronomy,
Nanjing University, for valuable discussions. Thanks also to Mr. C.
J. Pritchet for discussions and for his encouragement and
enlightenment to the first author. This work was supported by the
National Natural Science Foundation of China (grant No. 10873009)
and the National Basic Research Program of China (973 program) No.
2007CB815404.
\end{acknowledgements}

\appendix                  

\section{Observed Data \& Derived Jet Beaming Angles}

\center
 Table 1 Data from Observations \& Derived Data

\begin{tabular}{cccccc}
\hline ~~~~~~~~GRB~~~~~~~~&~~~~~~~~$E_{iso}$~~~~~~~~&~~~~~~~~t~~~~~~~~&~~~~~~~
~z~~~~~~~~&~~~~~~~~Category~~~~~~~~&~~~~~~~~$\theta_{j}(deg)$~~~~~~~~\\[0.5ex]
\hline
050709&$0.00069^{[1]}$&$10^{[1]}$&$0.16^{[1]}$&Type I$^{[2]}$&18.1957\\
060614&$0.021^{[2]}$&$1.3^{[4]}$&$0.13^{[4]}$&Type I$^{[2]}$&5.57885\\
970508&$0.061^{[2]}$&$25^{[4]}$&$0.835^{[4]}$&Type II$^{[2]}$&12.3366\\
980703&$0.72^{[2]}$&$2.49^{[4]}$&$0.966^{[4]}$&Type II$^{[2]}$&3.71802\\
990123&$22.9^{[2]}$&$1.8^{[4]}$&$1.6^{[4]}$&Type II$^{[2]}$&1.92367\\
990712&$0.0672^{[5]}$&$1.6^{[5]}$&$0.433^{[5]}$&Type II$^{[2]}$&4.77012\\
000418&$0.75137^{[6]}$&$25^{[6]}$&$0.1181^{[6]}$&Type II$^{[2]}$&8.54109\\
000926&$2.71^{[2]}$&$2.03^{[4]}$&$2.307^{[4]}$&Type II$^{[2]}$&2.40104\\
011211&$0.67234^{[6]}$&$1.77^{[6]}$&$2.14^{[6]}$&Type II$^{[2]}$&2.76815\\
020405&$1^{[2]}$&$2.74^{[4]}$&$0.689^{[4]}$&Type II$^{[2]}$&3.91555\\
020813&$6.6^{[2]}$&$0.46^{[4]}$&$1.254^{[4]}$&Type II$^{[2]}$&1.42145\\
000328&$3.61^{[5]}$&$0.8^{[5]}$&$1.52^{[5]}$&Type II$^{[2]}$&1.80902\\
030329&$0.166^{[5]}$&$0.5^{[5]}$&$0.169^{[5]}$&Type II$^{[2]}$&2.97281\\
041006&$0.83^{[5]}$&$0.16^{[5]}$&$0.716^{[5]}$&Type II$^{[2]}$&1.37314\\
050525A&$0.25^{[2]}$&$0.16^{[4]}$&$0.606^{[4]}$&Type II$^{[2]}$&1.63548\\
051221A&$0.024^{[7]}$&$4.1^{[4]}$&$0.5465^{[4]}$&short$^{[8]}$&7.50341\\
010921&$0.13611^{[6]}$&$33^{[6]}$&$0.4509^{[6]}$&long$^{[9]}$&13.5235\\
020124&$2.15^{[5]}$&$3^{[5]}$&$3.198^{[5]}$&long$^{[9]}$&2.61651\\
021004&$0.55601^{[6]}$&$7.6^{[6]}$&$2.332^{[6]}$&long$^{[9]}$&4.78799\\
030226&$0.67^{[5]}$&$0.84^{[5]}$&$1.986^{[5]}$&long$^{[9]}$&2.13395\\
030429&$0.173^{[5]}$&$1.77^{[5]}$&$2.656^{[5]}$&long$^{[9]}$&3.09816\\
050315&$0.49^{[10]}$&$2.6^{[4]}$&$1.95^{[4]}$&long$^{[11]}$&3.40529\\
050318&$0.22^{[3]}$&$0.12^{[4]}$&$1.44^{[4]}$&long$^{[11]}$&1.27531\\
050401&$5.323^{[12]}$&$0.06^{[12]}$&$2.9^{[12]}$&long$^{[11]}$&0.55384\\


050416A&$0.0083^{[5]}$&$1^{[5]}$&$0.653^{[5]}$&long$^{[11]}$&4.92335\\
050820A&$9.74^{[3]}$&$3.99^{[4]}$&$2.61^{[4]}$&long$^{[11]}$&2.55110\\
060124&$4.1^{[3]}$&$0.61^{[4]}$&$2.3^{[4]}$&long$^{[11]}$&1.45364\\
060206&$0.43^{[3]}$&$0.82^{[4]}$&$4.05^{[4]}$&long$^{[11]}$&1.83552\\
060210&$4.15^{[13]}$&$2.16^{[4]}$&$3.91^{[4]}$&long$^{[11]}$&2.00912\\
060526&$0.26^{[3]}$&$0.98^{[4]}$&$3.21^{[4]}$&long$^{[11]}$&2.23733\\
060605&$0.25^{[14]}$&$0.27^{[14]}$&$3.773^{[14]}$&long$^{[11]}$&1.32273\\
060814&$0.7^{[3]}$&$0.79^{[4]}$&$0.84^{[4]}$&long$^{[11]}$&2.48693\\
061121&$2.25^{[3]}$&$0.28^{[4]}$&$1.31^{[4]}$&long$^{[11]}$&1.33753\\
061126&$1.06^{[15]}$&$25.7^{[15]}$&$1.1588^{[15]}$&long$^{[11]}$&8.20785\\
070125&$10.6^{[16]}$&$3.8^{[16]}$&$1.547^{[16]}$&long$^{[16]}$&2.82483\\
071010A&$0.036^{[17]}$&$1^{[17]}$&$0.98^{[17]}$&long$^{[11]}$&3.83010\\
080319B&$13^{[18]}$&$0.032^{[18]}$&$0.937^{[18]}$&long$^{[11]}$&0.50876\\
970828&$2.1982^{[6]}$&$2.2^{[6]}$&$0.9578^{[6]}$&&3.09196\\
990510&$1.76349^{[6]}$&$1.2^{[6]}$&$1.6187^{[6]}$&&2.27045\\
990705&$2.55952^{[6]}$&$1^{[6]}$&$0.8424^{[6]}$&&2.30918\\
991216&$5.35369^{[6]}$&$1.2^{[6]}$&$1.02^{[6]}$&&2.17822\\
000301C&$0.43749^{[6]}$&$7.3^{[6]}$&$2.0335^{[6]}$&&5.03377\\
010222&$8.57841^{[6]}$&$0.93^{[6]}$&$1.4768^{[6]}$&&1.72899\\
\hline
\end{tabular}
\
 \tablerefs{0.86\textwidth}{[1] Fox et al. (2005);
 [2] Zhang et al. (2009);
 [3] Amati et al. (2008);
 [4] Liang et al. (2008);
 [5] Ghirlanda et al. (2007);
 [6] Bloom et al. (2008);
 [7] Soderberg et al. (2006);
 [8] Sakamoto et al. (2007);
 [9] P\'elangeon et al. (2008);
 [10] Amati L. (2007);
 [11] Evans et al. (2009);
 [12] Kamble et al. (2009);
 [13] Ghirlanda et al. (2008);
 [14] Ferrero et al. (2009);
 [15] Perley et al. (2007);
 [16] Chandra et al. (2008);
 [17] Covino et al. (2008);
 [18] Racusin et al. (2008).
 }

\label{lastpage}

\end{document}